\documentclass[%
 aip,
 amsmath,amssymb,
preprint,%
]{revtex4-2}

\usepackage{mathtools}
\usepackage{import}
\usepackage[pdftex]{graphicx}
\usepackage{dcolumn}
\usepackage{bm}
\usepackage{amsfonts}
\usepackage[utf8]{inputenc}
\usepackage[T1]{fontenc}
\usepackage{mathptmx}
\usepackage{etoolbox}
\usepackage{color}
\usepackage{multirow}
\usepackage{booktabs}
\makeatletter
\def\@email#1#2{%
 \endgroup
 \patchcmd{\titleblock@produce}
  {\frontmatter@RRAPformat}
  {\frontmatter@RRAPformat{\produce@RRAP{*#1\href{mailto:#2}{#2}}}\frontmatter@RRAPformat}
  {}{}
}%
\makeatother
\newtheorem{remark}{Remark}
\DeclareMathOperator*{\argmax}{argmax}
\begin{document}

\preprint{AIP/123-QED}

\title[Detecting Phase Synchronization in Latent Variable Subspace]{Detecting Phase Synchronization in Latent Variable Subspace:\\ Non-generating Partitions and Symbol Sequence Statistics}
\author{Henrique Carvalho de Castro}

\author{Luis Antonio Aguirre}

\affiliation{%
Graduate Program in Electrical Engineering,
Universidade Federal de Minas Gerais,\\
Av. Ant\^onio Carlos 6627, 31270-901, Belo Horizonte, MG, Brazil
}%

\email{hcastro@ufmg.br, aguirre@ufmg.br}
\date{\today}

\begin{abstract}
The detection of phase synchronization of coupled chaotic oscillators which are not phase-coherent is known to be a challenging task. In this work a method to detect and measure phase synchronization is presented. The procedure uses \textit{symbol sequence statistics} together with \textit{Principal Component Analysis} (PCA) and is applied in the phase synchronization analysis of pairs of coupled chaotic systems with different characteristics. Using PCA, we extract a $3$D space (called latent space) from the original $6$D space of the coupled oscillators. When  the oscillators are in complete synchronization, the latent space represents the dynamics of an isolated oscillator. However, as synchronization deteriorates, the latent space becomes increasingly disorganized, although it does retain some level of organization during phase synchronization. A $2$D Poincar\'e-type section is defined in the latent space and the corresponding $1$D map is used to define a non-generating partition such that an arbitrary symbol sequence is forbidden for any synchronized regime. It is shown that the probability of occurrence of such a symbol sequence is closely related to the quality of phase synchronization. The procedure does not require a phase definition or complicated partitioning algorithms, which is performed by a simple threshold-crossing technique. This method requires data from different levels of synchronization to be able to determine the required non-generating partition. 
\end{abstract}

\maketitle

\begin{quotation}
Detecting and measuring phase synchronization between chaotic oscillators are not trivial tasks. Many approaches involve creating a phase model, or estimating the phase. 
An approach is here introduced that does not require a proper definition of phase. It is based on the monitoring of the so called \textit{latent variables}, or principal components, derived from a linear transformation of the original variables. As the phase synchronization deteriorates, the latent space becomes disorganized and this phenomenon can be characterized through coarse grained codification of the trajectories. Such codification is performed in such a way that a specific symbol sequence is forbidden during a scenario of good quality phase synchronization. The mere presence of this word should indicate divergence in the oscillators phase. The technique is demonstrated using different schemes of coupling benchmark simulated systems, and compared with the traditional phase coherence measurement.
\end{quotation}

\section{\label{sec:intro}Introduction}

Phase synchronization is a phenomenon that arises from the interaction between nonidentical autonomous chaotic oscillators, in which there is a locking in their phases whereas their amplitudes remain chaotic and non-correlated \cite{pik_eal/97b}. A common approach to address problems involving phase synchronization is to reduce complex oscillators to a phase model\cite{nak/16}. In some situations, it is more interesting to describe the synchronization phenomenon without simplifications or reductions, so that different dynamical regimes such as complete, phase, lag, intermittent, and generalized synchronization can also be defined\cite{boc_eal/02}. Both approaches are linked to a common challenge: to define a proper phase variable that can be applied to a wide class of oscillators\cite{fre_eal/18}. Thus, avoiding a proper definition of a phase variable, or model, during phase synchronization analysis should simplify and generalize the process.

One such a method relies on the \textit{(multivariate) singular spectrum analysis} (M-SSA) \cite{Vautard_etal1992,Muller_etall2005,Feliks_etall2010,Groth_Ghil2011,Portes_Aguirre2016}. \citeauthor{Groth_Ghil2011}\cite{Groth_Ghil2011} show that the M-SSA is able to automatically identify oscillatory modes and detect cluster synchronization in large systems of coupled oscillators. Such oscillatory modes correspond to shared dynamical behavior between the clustered systems. It is based on previous work done by \citeauthor{Vautard_Ghil1989}\cite{Vautard_Ghil1989}, who show that when a vigorous oscillation is present, a pair of nearly equal eigenvalues stands out of the singular spectrum, and that the associated eigenvectors and \textit{principal components} (PCs) are in quadrature. That is, the pair of eigenvectors that correspond to the same period are the data-adaptive equivalent of sine-and-cosine pairs in Fourier analysis.

The M-SSA is an extension of the the \textit{principal component analysis} (PCA), and it has been used  in nonlinear dynamics \cite{Broomhead_King1986,Fraedrich1986}, and in several fields of the geosciences and other disciplines \cite{Ghil_etal2002}. The advantage over PCA is that M-SSA is able to capture dynamical behavior, while PCA is a static `model' that assumes time independence between data samples. This allows for the reconstruction of a robust  ``skeleton'' of the dynamical structure underlying the data.\cite{Vautard_etal1992,Groth_Ghil2011} The central idea of PCA is to reduce the dimensionality of a data set consisting of many interrelated variables while retaining as much as possible of the variance present in the data set. For a data set composed of $D$ variables, one looks for a few ($\ll D$) derived variables that account for most of the data variability. Such derived variables are called PCs, or \textit{latent variables}, and are defined as a linear transformation of the original variables into a new set of variables that are uncorrelated to each other.

The aforementioned technique to detect phase synchronization with no proper definition of phase is based on the observation of the singular spectrum, i.e. the eigenvalues yielded from the M-SSA. However, its application is not straightforward.  Due to the degeneracy of the eigenvectors, the ability of the ``original'' M-SSA to identify the formation of clusters of synchronized oscillators decreases drastically as the number of oscillators increases\cite{Groth_Ghil2011}. To overcome this problem varimax rotations of the M-SSA eigenvectors are proposed\cite{Groth_Ghil2011,Portes_Aguirre2016}.

In this paper we introduce a new angle to address phase synchronization by monitoring the latent variables instead of the singular spectrum. We define the \textit{latent variable subspace} whose topology is highly dependent on the synchronization level. Our first approach is to characterize the latent space topology using coarse grained observations coded into symbol sequences. We show that as the phase synchronization onset deteriorates the latent space becomes disorganized, and this phenomenon can be measured through the observation of a specific symbol sequence (the forbidden word) previously defined.

The objective of this work is threefold: (i) given a pair of coupled chaotic oscillators,  analyze the latent variable subspace (extracted using PCA) during the transition to the onset of phase and complete synchronization; (ii) by doing so, assess the use of PCA in discriminating different synchronization regimes; (iii) propose a method to detect phase synchronization regime in the latent subspace. 

The paper is organized as follows. Section~\ref{sec:back} introduces basic concepts necessary for the work. Section~\ref{sec:problem} presents the problem and methodology. Results, discussion and exemplifications are given in Section~\ref{sec:results}. The main conclusions of the paper are provided in Section~\ref{sec:conclusion}.

\section{Background}
\label{sec:back}

\subsection{Phase Synchronization}
\label{ssec:sync}

Complete synchronization (CS) of identical oscillators coupled using all variables was
investigated in the pioneering paper.\cite{fuj_yam/83} The fact that CS could be achieved
by coupling a single variable was pointed out a decade later.\cite{kap/94} 
Another important move in the field was to realize that there is a weaker form of synchronization
that has been named phase synchronization (PS).\cite{ros_eal/96a} It soon became
evident that this type of synchronization was of great importance in a number of
fields.\cite{piq/11,dor_bul/14,pec_car/15,ero_eal/17}

Sometimes it is possible to find a plane $(x,\,y)$ on which there is a well defined center 
of rotation.\cite{let/20} In such cases the phase can be defined as:\cite{pik_eal/97b}
\begin{equation} 
	\label{eq:phase_atan}
	\phi(t) \coloneqq {\rm arctan2} \left(y,x\right),
\end{equation}
where $\phi(t) \in [-\pi; \pi)$. When the aforementioned rotation plane is not readily obtained
by projection of the oscillator coordinates as for the Lorenz, ``funnel'' R\"{o}ssler, Cord \cite{agu_let/11} and Li \cite{li/08} 
attractors, computing the phase with (\ref{eq:phase_atan}) is not generally possible.\cite{fre_eal/18} For some
of those cases specific transformations have been used, e.g.  for Lorenz \cite{pik_eal/97a} and 
funnel R\"{o}ssler.\cite{che_eal/01}

A phase definition based on the concept of curvature is:\cite{kur_eal/06}
\begin{equation} 
	\label{eq:phase_curv}
	\phi(t) \coloneqq {\rm arctan2} \left(\dot{y},\dot{x}\right) ,
\end{equation}
which is the definition that will be used to validate our results because it is well suited for 
a variety of cases including the spiral and funnel R\"{o}ssler attractors.

\citeauthor{Mormann_etal2000}\cite{Mormann_etal2000} propose the 
use of \textit{mean phase coherence of an angular
		distribution} as a measure of synchronization. It is defined as:
	\begin{equation}
		\label{eq:meanPhaseCohe}
		R = \left| \frac{1}{N}\sum_{k=0}^{N-1}{\rm e}^{i\phi(kT_{\rm s})}\right|,
\end{equation}
where $1/T_{\rm s}$ is the sampling rate of the discrete time
series.

\subsection{The Latent Variable Subspace}
\label{ssec:pca}

Consider, for the moment, a set of $D$ vectors 
\[
\bm{X} = \{\bm{x}_i\},\, i=1,\dots,D,
\] 
where $\bm{X}\in\mathbb{R}^{N\times D}$ and $\bm{x}_i\in\mathbb{R}^N$, with covariance 
matrix \mbox{$\bm{S}=\textbf{X}^T\bm{X}/N$}. Denote by $\bm{z}$ a linear combination of vectors $\bm{x}_i$ 
that retain most of the relevant information in $\bm{X}$ and assume that the $N$ observations 
are independent. Then
\begin{equation}
	\bm{z}=\bm{X}\bm{u},
\end{equation}

\noindent
where $\bm{z}\in\mathbb{R}^{N}$ and $\bm{u}\in\mathbb{R}^{D}$. To retain as much of the variation 
in $\bm{X}$ as possible, one must maximize the variance of $\bm{z}$ choosing optimal weights in $\bm{u}$, 
requiring that the norm, i.e. the sum-of-squared values, is one, that is
\begin{equation}
	\label{eq:pca:maximization}
	\argmax_{||\bm{u}||=1} {\rm var}(\bm{z}) \equiv \argmax_{||\bm{u}||=1}\, (\bm{z}^T\bm{z}) \equiv \argmax_{||\bm{u}||=1}\, (\bm{u}^T\bm{X}^T\bm{X}\bm{u}),
\end{equation}

\noindent
where it is assumed that $\bm{X}$ is mean-centered, and $(\cdot)^T$ stands for transposition. 
The last equivalence in \eqref{eq:pca:maximization} is a standard problem in linear algebra and 
the optimal $\bm{u}$ is the eigenvector of $\bm{S}$ that corresponds to the largest eigenvalue.

Call $\{\bm{u}_1 \ldots \bm{u}_D\}$ the eigenvectors of $\bm{S}$ ordered according to the
corresponding eigenvalues $\lambda_1 > \ldots > \lambda_{D}>0$ and form 
the transformation matrix $\bm{U}=[\bm{u}_1 \ldots \bm{u}_p]\in\mathbb{R}^{D\times p}$ such that: 
\begin{equation}
	\label{eq:PCA}
	\bm{Z} = \bm{XU},
\end{equation}

\noindent
where $\bm{Z}=[\bm{z}_1 \ldots \bm{z}_p]\in\mathbb{R}^{N\times p}$, and $\bm{z}_k$ is the $k$th 
\textit{principal component} (PC) or \textit{latent variable}, of $\bm{X}$. The transformation of $\bm{X}$ into $\bm{Z}$
in \eqref{eq:PCA} is central to PCA.

The problem of determining the $p$ latent variables that best represent the data $\bm{X}$ is treated elsewhere \cite{Cattel1966,Farmer1971,krzanowski_Kline1995,Bro_etal2008,Bro_Smilde2014}
and will not be discussed here. 

For the sake of presentation let us rewrite matrix $\bm{X}$ as $\bm{X}_1^N$ and build
an augmented trajectory matrix $\bm{X}_{\rm a}$ by including shifted 
versions of the variables in $\bm{X}_1^N$, such that:
\begin{equation*}
	\bm{X}_{\rm a} = \{ \bm{X}_1^{N{-}M{+}1} \bm{X}_2^{N{-}M{+}2}  \ldots \bm{X}_M^{N}\},
\end{equation*}

\noindent
where  $\bm{X}_{\rm a}\in\mathbb{R}^{N-M+1\times DM}$ and $M$ is the maximum delay of the shifted 
versions of each variable in $\bm{X}$. The M-SSA consists in applying the traditional PCA on $\bm{X}_a$. 

The latent variable subspace (or \textit{latent space} for short) is the subspace spanned by the first $p$ 
eigenvectors of the covariance matrix of $\bm{X}$ (or $\bm{X}_{\rm a}$), which are the first $p$ columns of 
$\bm{U}$. Such columns span a $p$-dimensional subspace that explains most of the variation of 
$\bm{X}$ (or $\bm{X}_{\rm a}$).

\subsection{Singular Value Decomposition}

The \textit{singular value decomposition} (SVD) provides a computationally efficient method of 
finding PCs. Briefly, the SVD of a matrix $\bm{X}$ with rank $r$ is:
\begin{equation}
	\label{eq:svd}
	\bm{X} = \bm{U^\prime LA}^T,
\end{equation}

\noindent
where $\bm{U^\prime}\in\mathbb{R}^{N \times r}$, $\bm{A\in\mathbb{R}^p \times r}$ are 
orthonormal, that is, ${\bm{U^\prime}}^T\bm{U^\prime} = \bm{I}_r$, $\bm{A}^T\bm{A} = \bm{I}_r$
and $\bm{L}{=}{\rm diag}\{ \sqrt{\lambda_1} \ldots \sqrt{\lambda_r} \} \in\mathbb{R}^{r \times r}$.
$\bm{U^\prime}$ in \eqref{eq:svd} should not be confused with $\bm{U}$ in \eqref{eq:PCA}.
In \eqref{eq:svd}, $\bm{A}$ are the eigenvectors of $\bm{X}^T\bm{X}$, in other words, the
columns of $\bm{A}$ are the eigenvectors of $\bm{S}$ and the diagonal elements of $\bm{L}$
are the standard deviations of the respective principal components of $\bm{X}$, obtained from \eqref{eq:PCA} with $\bm{U}=\bm{A}$.\cite{Jolliffe2002}

\subsection{Symbol Sequences}
\label{ssec:symb}

In many situations it is important to quantify dynamic complexity. 
Several approaches for attaining this are based on information 
theory\cite{Cover_Thomas1991} and require sequences of \textit{coarse-grained} observables, 
which typically have to be generated from continuous-valued discrete time series. 

Any trajectory of a dynamical system can be encoded as an infinite sequence of symbols. 
For this it is necessary to partition the state space into disjoint subsets, each represented 
by a unique symbol from a finite alphabet. Such a partition provides an equivalent description 
of the dynamics in the continuous state space when the assignment of symbol sequences 
to trajectories is unique, that is, every infinite symbol sequence corresponds to a unique point 
in state space. This particular kind of partition is called \textit{generating} and it preserves all 
deterministic dynamical information in the symbolic representation. The construction of generating 
partitions is not obvious for systems of dimension greater than one.
Several approaches have been proposed to partition bi-dimensional and higher-dimensional systems\cite{Grassberger_Kantz1985,Davidchack_etal2000,Plumecoq_Lefranc2000_p1,Plumecoq_Lefranc2000_p2,Hirata_etal2004,Patil_Cusumano2018}, but no general method exists. The theory on \textit{symbolic dynamics} is formalized in Ref.\,(\onlinecite{Collet_Eckmann1980}), where it is shown that a complete description of the behavior of a dynamical system can be encoded in a symbol sequence using a generating partition.  It is a powerful tool to characterize non linear dynamics with applications in several fields, as hydrodynamics \cite{Godelle_Letellier2000} and cardiodynamics \cite{Fresnel_etal2015}.

In this paper,  \textit{non-generating} partitions of the state space are used 
and hence symbolic dynamics are not formalized here. However, some notations will be useful.

Consider a continuous dynamical system vector field \mbox{$f:M\rightarrow M$}. Let $\mathcal{A}=\{0,1,\dots,q-1\}$ be an alphabet of size $q$, and $\mathbf{\mathcal{M}}=\{M_0,M_1,...,M_{q-1}\}$ be a partition which divides the sate space into $q$ disjoint sets. Define a coding function $s\,:\,M\to\mathcal{A}$ as:
\begin{equation*}
	s(\bm{x}) = \alpha\,\iff\, \bm{x} \in M_\alpha,
\end{equation*}

\noindent
where $\alpha$ is a symbol.
The set of all infinite symbol sequences that system $f$ can realize is denoted by 
$\Sigma$ and is called \textit{shift space}, and the set of all realizable $L$-block words 
is denoted $\Sigma^{[L]}$ and is called \textit{higher-block shift space}. It is important to note that the latter $L$-blocks are extracted from the symbol sequence by shifting and overlapping, that is
\begin{equation*}
	\mathrlap{\overbrace{\phantom{0\,\,1\,\,0\,\,0\,\,1\,\,}}^{s_1}}0\,\,\mathrlap{\underbrace{\phantom{1\,\,0\,\,0\,\,1\,\,1\,\,}}_{s_2}}1\,\,0\,\,0\,\,1\,\,1\,\,0\,\,1\,\,1\,\,1\,\,0\,\,1\,\dots
\end{equation*}
The $5$-block words $s_1=01001$ and $s_2=10011$ are drawn from the first $6$ symbols of the sequence.

In this work, a non-generating partition $\mathcal{M}$ is arbitrarily chosen such  
that some $L$-block word $s^{[L]}(\bm{x})\in\Sigma^{[L]}$ is only allowed when a specific situation happens. 
For instance, such a word may be used as an indication of phase slipping between coupled oscillators.

\section{Detecting Phase Synchronization}
\label{sec:problem}

\subsection{Problem Statement}

Let $\bm{x}_i^{\rm u}(t){\in}\mathbb{R}^3, i{=}1,2$ be the trajectories of two 
nonidentical 3D {\it un}coupled oscillators with highly dissipative chaotic dynamics
sampled regularly with sampling time $T_{\rm s}$. 
After coupling both oscillators the trajectories become $\bm{x}_i(t){\in}\mathbb{R}^3, i{=}1,2$.
The problem addressed in this paper is to determine from $\bm{x}_i(t), i{=}1,2$
whether the oscillators are phase synchronized using as additional information
the uncoupled trajectories $\bm{x}_i^{\rm u}(t), i{=}1,2$.

\subsection{Methodology}
\label{subsec:methodology}
For the sake of clarity, the procedure is described as a sequence of steps.

\begin{enumerate}
	\item  Construct the trajectory matrix $\bm{X}^{\rm u} \in\mathbb{R}^{N\times 6}$ 
	by stacking the trajectories of both uncoupled oscillators:
	\begin{equation}
		\label{eq:augmentedTrajectory}
		\bm{X}^{\rm u} = \left[ \begin{tabular}{cc}
			$\bm{x}_1^{\rm u}(t)$ & $\bm{x}_2^{\rm u}(t)$
		\end{tabular} \right]_{t=0,\,...,\,(N{-}1)T_{\rm s}}.
	\end{equation} 
	
	\item
	Construct a trajectory matrix  $\bm{X}^{\rm i} \in\mathbb{R}^{N\times 6}$ to
	represent what identical synchronization would look like:
	\begin{equation}
		\label{eq:augmentedTrajectory_i}
		\bm{X}^{\rm i} = \left[ \begin{tabular}{cc}
			$\bm{x}_i^{\rm u}(t)$ & $\bm{x}_i^{\rm u}(t)$
		\end{tabular}
		 \right]_{t=0,\,...,\,(N{-}1)T_{\rm s}},
	\end{equation} 
	
	\noindent
	where $i$ can be chosen to be either 1 or 2.

	\item
	Using \eqref{eq:svd}, perform the SVD of $\bm{X}^{\rm u}$ and $\bm{X}^{\rm i}$.
	
	\item
	Compute the PCs of $\bm{X}^{\rm u}$ and $\bm{X}^{\rm i}$ as: 
	$\bm{Z}^{\rm u} {=} \bm{X^{\rm u}A^{\rm u}}$ and $\bm{Z}^{\rm i} {=} \bm{X^{\rm i}A^{\rm i}}$ (see Eq.\,\ref{eq:PCA}).	This is the ordinary PCA algorithm. 
	
	\item
	Define the latent variable subspace as
	\begin{equation}
		\label{eq:latentSpace}
		\bm{Z}_{1:3}^{\rm i} = \left[\begin{tabular}{ccc}
			$x^\prime(t)$ & $y^\prime(t)$ & $z^\prime(t)$
		\end{tabular} \right]_{t=0,\,...,\,(N{-}1)T_{\rm s}},
	\end{equation}
	
	\noindent
	where $\bm{Z}_{1:3}^{\rm i}\in\mathbb{R}^{N\times3}$ is composed
	by the first three columns of $\bm{Z}^{\rm i}$, and $(x^\prime(t),y^\prime(t),z^\prime(t))$ 
	are the first, second and third PCs, respectively, of $\bm{X}^{\rm i}$. Proceed likewise for $\bm{Z}^{\rm u}$.
	
	\item
	Graphically represent $\bm{Z}_{1:3}^{\rm i}$ in the latent space and construct a 2D section $\mathcal{P}$
	as if it were a Poincar\'e section.
	
	\item
	Using the data on $\mathcal{P}$ find a 1D first return map \mbox{$P\,:\,z[k{-}1] \mapsto z[k]$},
	where $z[k]{\in}[\,z_{\rm min}~z_{\rm max}\,]$ is one of the PCs.
	
	\item
	Call $z_{\rm th}$ a threshold value based on which the following coding
	function is defined
	\begin{align}
		\label{eq:partition}
		\mathcal{I}(z) = 
		\begin{cases}
			0\,,\\
			1\,,
		\end{cases}
		\begin{split}
			\text{if}\;\; z [k] < z_{\rm th},\\
			\text{otherwise}
		\end{split}
	\end{align}
	for all $k$ over the data on $\mathcal{P}$. Vary $z_{\rm th}$ within $[z_{\rm min}~z_{\rm max}]$
	and for each threshold value, code the data $ z [k]$ according to \eqref{eq:partition}, for both latent variable subspaces $\bm{Z}^{\rm i}_{1:3}$ and $\bm{Z}^{\rm u}_{1:3}$.
	
	\item
	In order to define a partition: i)~choose $q$, the number of symbols in a word; for each value of $z_{\rm th}$ 
	within the range defined in the previous step ii)~construct the higher block shift spaces $\Sigma_{\bm{Z}^{\rm i}}^{[q]}$ and $\Sigma_{\bm{Z}^{\rm u}}^{[q]}$  and iii)~compute the probability of the word with $q$ ones 
	\mbox{$P(\Sigma^{[q]}=1\ldots1|z_{\rm th})$},
	which is a ``forbidden word'' in the sense that it should {\it not} happen in case
	of complete synchronization. Hence choose $z_{\rm th}$ and $q$ such that
	\mbox{$P\big(\Sigma^{[q]}_{\bm{Z}^{\rm i}}=1\ldots1|z_{\rm th}\big){<}\epsilon$}, where $0{<}\epsilon {\ll}1$, and \mbox{$P\big(\Sigma^{[q]}_{\bm{Z}^{\rm u}}=1\ldots1|z_{\rm th}\big)$} is largest.
	
	\item 
	Construct the trajectory matrix $\bm{X} \in\mathbb{R}^{N\times 6}$ with the test data
	by stacking the trajectories of both coupled oscillators:
	\begin{equation}
		\label{eq:augmentedTrajectory}
		\bm{X} = \left[\begin{tabular}{cc}
			$\bm{x}_1(t)$ &
			$\bm{x}_2(t)$
		\end{tabular}
		\right]_{t=0,\,...,\, NT_{\rm s}}.
	\end{equation} 
	Repeat steps 3 through 7 for $\bm{X}$. Step~8
	should be repeated only for the the values of $z_{\rm th}$ and $q$ chosen in Step~9.
	
	\item
	If the forbidden word did {\it not} happen, then the oscillators are phase syncrhonized
	with high quality. The quality of the synchronization degrades as the number of
	occurrences of the forbidden word increases. 
\end{enumerate}

\begin{remark}
	Matrix $\bm{X}^{\rm i}$ represents the hypothetical situation of identical or complete synchronization.
	Hence one oscillator is duplicated. In the case of nonidentical
	oscillators, in principle any of them can be used. More on this later.
\end{remark}

\begin{remark}
	The choice of how many components should be kept to define the latent space 
	is based on the dimensionality of the oscillators.
\end{remark}

\begin{remark}
	The problem of phase synchronization detection is here handled by translating a section of the latent space into symbolic sequences using a \textit{non-generating} partition such that  one (or some) of the $L$-block words in $s^{[L]}(z)$ is forbidden in the phase synchronized regime and allowed otherwise.
\end{remark}

\begin{remark}
In step~$9$, we propose that $P\big(\Sigma^{[q]}_{\bm{Z}^u}=1\ldots1|z_{\rm th}\big)$ be maximum. It is not clear yet what should be an acceptable value for $P\big(\Sigma^{[q]}_{\bm{Z}^u}=1\ldots1|z_{\rm th}\big)$. In this paper, we work with the hypothesis that it should be sufficiently large as to ``dominate" the dynamics during the non synchronized state. That is, the probability of occurrence of the forbidden word must be much greater than all others in such a situation.
\end{remark}

\section{\label{sec:results}Results and Discussion}
Consider a pair of possibly non-identical coupled R\"ossler oscillators:
\begin{equation}
	\label{eq:rossler}
	\bm{x}_{1,2} = 
	\begin{cases}
		\Dot{x}_{1,2} = -\omega_{1,2}\,y_{1,2}-z_{1,2}+\kappa\,(x_{2,1}-x_{1,2})\\
		\Dot{y}_{1,2} = \omega_{1,2}\,x_{1,2}+ a\,y_{1,2}\\
		\Dot{z}_{1,2} = b+z_{1,2}(x_{1,2}-c),
	\end{cases}
\end{equation}
where $a\in\{0.165, 0.280\}$, $b=0.2$, $c=10$, and $\kappa$  is the coupling strength. Let $\omega_{1,2}=\omega_0 \pm \Delta$, in which $\omega_0=0.97$, and $\Delta\in[\,0.00, 0.03\,]$. The parameter $\Delta$ is responsible for a small mismatch of the oscillators natural frequencies. And a pair of possibly non-identical Lorenz oscillators:
\begin{equation}
	\label{eq:lorenz}
	\bm{x}_{1,2} = 
	\begin{cases}
		\Dot{x}_{1,2} = \sigma(w\,y_{1,2}-x_{1,2}) +\kappa\,(x_{2,1}-x_{1,2})\\
		\Dot{y}_{1,2} = r\,x_{1,2}-w\,y_{1,2}-x_{1,2}\,z_{1,2}\\
		\Dot{z}_{1,2} = x_{1,2}\,w\,y_{1,2} - b\,z_{1,2},
	\end{cases}
\end{equation}

\noindent
where $\sigma=10$, $b=8/3$, $r=28$, and $\kappa$ is the coupling strength. The parameter $w$ is responsible for a small mismatch in the evolution of the oscillators $y$ variable, which should influence their natural frequencies.

In the following, we analyze the \textit{latent variable subspace} extracted from the coupled systems for different values of the coupling strength ($\kappa$) and of the natural frequency mismatch ($\Delta$ and $w$), comprising the regimes of complete synchronization, phase synchronization, and no synchronization. Furthermore, by changing $a$ the R\"ossler oscillator can operate in both spiral and funnel regimes.


The numerical setup is the following.
\paragraph{Partitioning rule.} Steps~1 through 9 described in Section~\ref{subsec:methodology} are performed by
considering a pair of uncoupled oscillators, as defined in \eqref{eq:rossler} and \eqref{eq:lorenz}, for $\kappa{=}0.00$. The systems are integrated numerically, using a fourth-order Runge-Kutta method with integration step $\delta t{=}0.01$ and with the same sampling time, that is, $T_{\rm s} = \delta t$, for $30$ Monte Carlo simulations varying the values of $\Delta$, $w$ and \textit{initial states} within ranges that will be described later. The transient regime is discarded by cutting off the first half of the data, and keeping the remaining $N{=}7.5 {\times} 10^5$ samples.

\paragraph{Statistics and validation of symbol sequence.} With the data resulting from simulations with $\kappa>0$ and the partitioning rules previously defined, the complement of the forbidden word probabilities are compared with the mean phase coherence, Eq.~\eqref{eq:meanPhaseCohe}.\\

The simulations are divided into two groups according to the topological similarity between the coupled oscillators as follows:
\begin{itemize}
	\item Similar systems
	\begin{itemize}
		\item (R\"ossler) spiral-spiral
		\item (R\"ossler) funnel-funnel
		\item Lorenz-Lorenz
	\end{itemize}
	\item Different systems
	\begin{itemize}
		\item (R\"ossler) spiral-funnel
	\end{itemize}
\end{itemize}

For the sake of clarity and compactness, they are all presented together.

\subsection{Partitioning rules}

Following Steps~1 through 5 presented in Section~\ref{subsec:methodology},  the latent variable subspaces 
$\bm{Z}^{\rm i}_{1:3}$ and $\bm{Z}^{\rm u}_{1:3}$ were found for all coupling schemes, which correspond 
to the cases of hypothetical identical synchronization and uncoupled oscillators, respectively. 
The 2D sections $\mathcal{P}_{\rm spiral}$, $\mathcal{P}_{\rm funnel}$, and $\mathcal{P}_{\rm lorenz}$ are defined based on the traditional Poincar\'e section for the respective oscillators (step 6): 
\begin{align}
	\label{eq:section}
		&\nonumber\mathcal{P}_{\rm spiral} = \{(x^\prime,z^\prime)\in\mathbb{R}^2\;|\;y^\prime=\overline{y}^\prime_0,\;x^\prime<\overline{x}^\prime_0\},\\
		&\mathcal{P}_{\rm funnel} = \{(x^\prime,y^\prime)\in\mathbb{R}^2\;|\;z^\prime=\overline{z}^\prime_0,\;y^\prime<\overline{y}^\prime_0\}, \\
		&\nonumber\mathcal{P}_{\rm lorenz} = \{(x^\prime,z^\prime)\in\mathbb{R}^2\;|\;y^\prime{=}\overline{y}^\prime_1,\;x^\prime{<}\overline{x}^\prime_1\;\cup\;y^\prime{=}\overline{y}^\prime_2,\;x^\prime{>}\overline{x}^\prime_2\},
\end{align}

\noindent
where $\overline{x}^\prime_i$, $\overline{y}^\prime_i$ and $\overline{z}^\prime_i$ are the first, second and third components, respectively, of the PCA transformation of the augmented fixed point
\begin{equation*}
	\label{eq:augmentedFixed}
	\overline{\bm{X}}_i = \left[\begin{tabular}{cc}
		$\overline{\bm{x}}_{i,1}$ &
		$\overline{\bm{x}}_{i,2}$
	\end{tabular}
	\right].
\end{equation*}

The fixed point of interest from the R\"ossler oscillator is given by
\begin{align}
	\nonumber\overline{\bm{x}}_0 {=} \left( \frac{c{-}\sqrt{c^2 {-} 4\,a\,b}}{2},\frac{{-}c{+}\sqrt{c^2 {-} 4\,a\,b}}{2\,a},\frac{c{-}\sqrt{c^2 {-} 4\,a\,b}}{2\,a}\right).
\end{align}
And the ones from the Lorenz oscillator are
\begin{align}
	&\nonumber\overline{\bm{x}}_1 {=} \left(\sqrt{b\,(r{-}1)},\sqrt{b\,(r{-}1)},r{-}1\right),\\
	&\nonumber\overline{\bm{x}}_2 {=} \left({-}\sqrt{b\,(r{-}1)},{-}\sqrt{b\,(r{-}1)},r{-}1\right).
\end{align}

\begin{figure*}
	\centering
	\includegraphics[width=\linewidth]{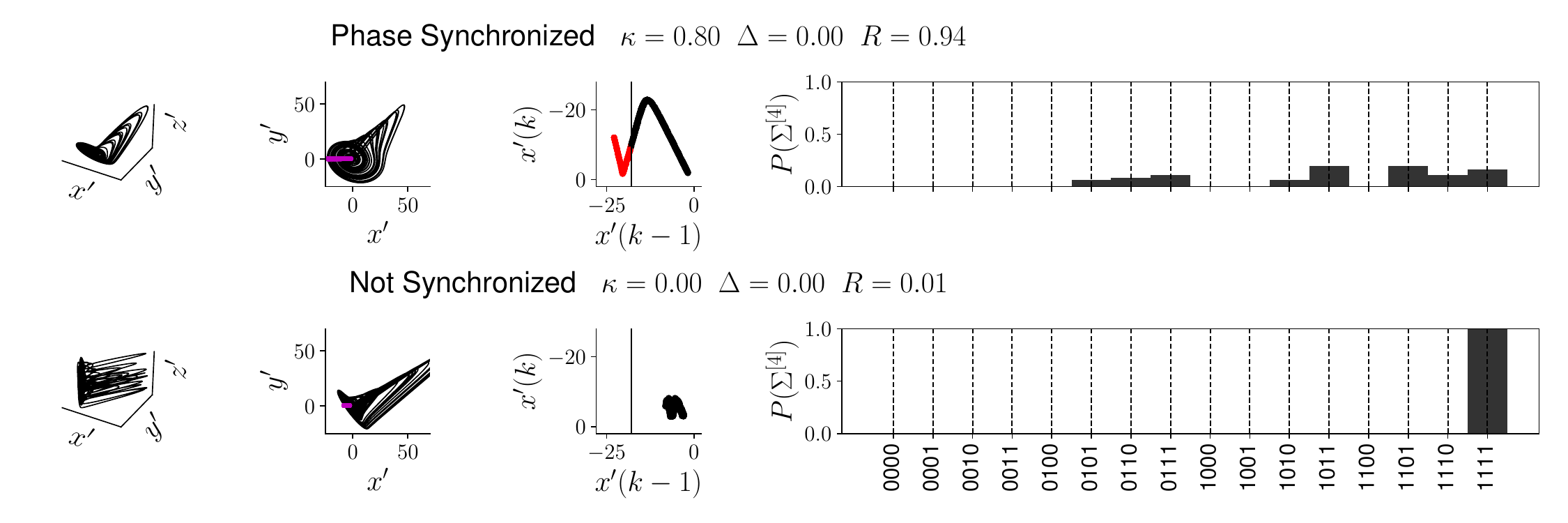}
	\caption{Latent spaces for the cases of phase synchronization (top) and no synchronization (bottom) between a pair of R\"ossler oscillators in different regimes: spiral and funnel. The first column presents the $3$D latent space (first three principal components). The second column presents their projection onto the $2$D plane together with the section \mbox{$\mathcal{P}_{\rm spiral}$ (\textcolor{magenta}{\hspace{0.1ex}\rule[0.297ex]{0.4cm}{1.5pt}\hspace{0.1ex}})}. The third column presents the first return map together with the chosen non-generating partition over $\mathcal{A} = \{0 (\textcolor{red}{\bullet}),\,1 (\textcolor{black}{\bullet})\}$. The histograms on last column represent the symbol sequence statistics for the given non-generating partitions over $\mathcal{A}$ in the higher block shift space $\Sigma^{[4]}$. The $4$-block words are in lexicographic order. The projections shown in the second column are built using only a part of the data used in the corresponding first return maps. The titles present the respective coupling strength ($\kappa$), natural frequency mismatch ($\Delta$), and mean phase coherence ($R$).}
	\label{fig:spiralFunnel}
\end{figure*}

Notice that the proposed sections in \eqref{eq:section} do not take into account the direction of the flow, i.e. such sections sample the trajectories in both crossing directions. Further, they limit the range of the $x^\prime$ and $y^\prime$ variables to make them general for the whole synchronization spectrum (more details in further examples). High dissipation allows the construction of 1D first return maps (step 7) \mbox{$P_{\rm spiral}:x^\prime[k{-}1]\mapsto x^\prime[k]$}, \mbox{$P_{\rm funnel}:y^\prime[k{-}1]\mapsto y^\prime[k]$}, and \mbox{$P_{\rm lorenz}:\big|x^\prime[k{-}1]\big|\mapsto \big|x^\prime[k]\big|$}. For the (R\"ossler) spiral-funnel coupling scheme, the section and mapping definitions are similar to the (R\"ossler) spiral-spiral coupling scheme despite during phase synchronization regime its topology resembles the funnel one (see first return map in Fig.~\ref{fig:spiralFunnel}).

Following Step~8, the following ranges for the threshold were chosen and used in coding the data 
for the hypothetical identical synchronization and for the uncoupled cases: \mbox{${x^\prime_{\rm th}}^{\rm spiral}\in[{-}22~~0]$}, \mbox{${y^\prime_{\rm th}}^{\rm funnel}\in[{-}40~~0]$}, and \mbox{${x^\prime_{\rm th}}^{\rm lorenz}\in[0~~45]$}.  
For the sake of investigation the two possible hypothetical latent spaces were considered 
$\bm{Z}^{\rm i}=\bm{Z}^1$ and $\bm{Z}^{\rm i}=\bm{Z}^2$, built from the duplication of $\bm{x}_1(t)$ and $\bm{x}_2(t)$, respectively. The unsynchronized (uncoupled) latent space $\bm{Z}^{\rm u}$ was also coded, as required.

\begin{figure}[h!]
	\includegraphics[width=0.5\linewidth]{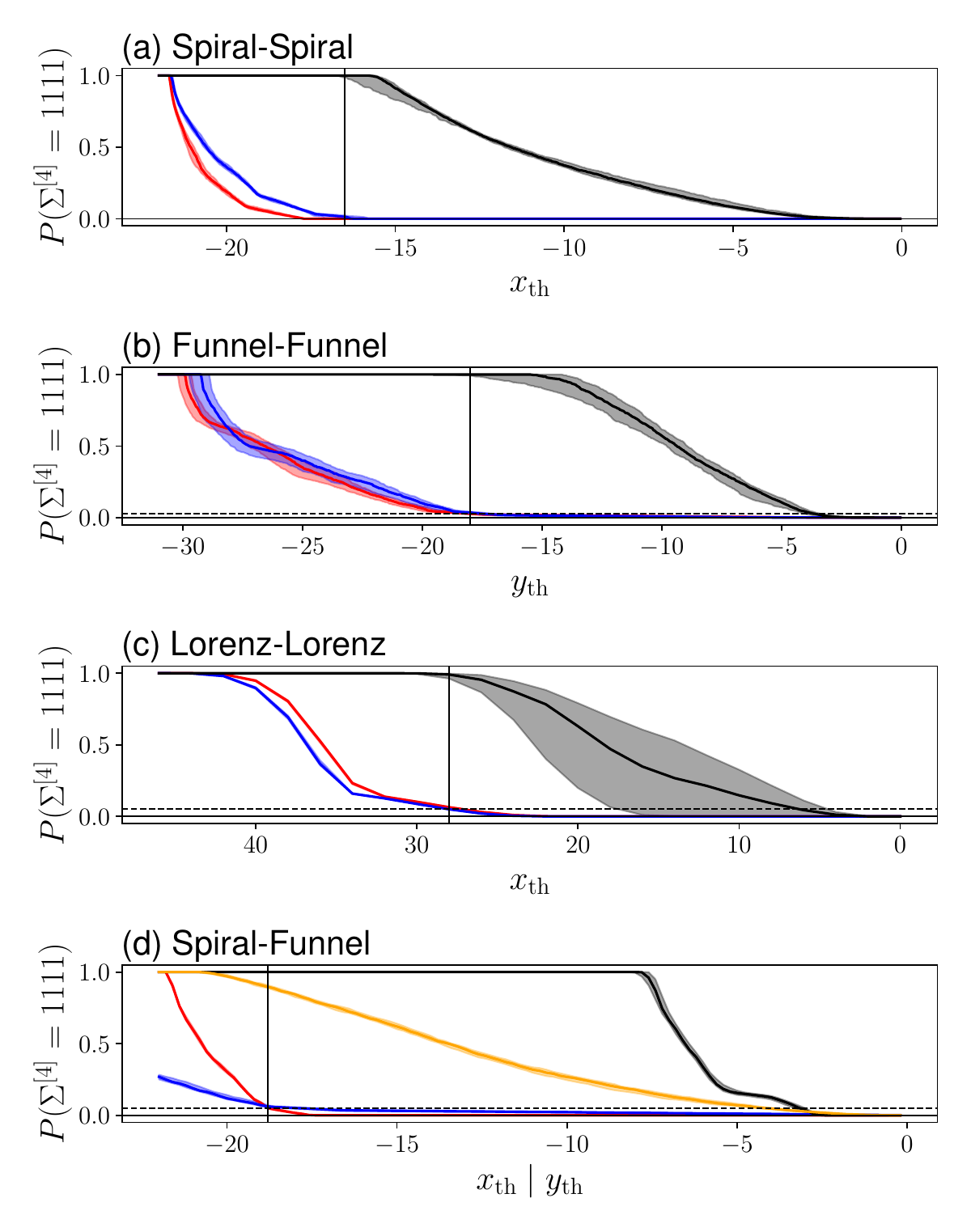}
	\caption{Four-block forbidden word probabilities vs. threshold coordinate. The vertical line indicates the chosen threshold value: (a)~$x_{\rm th}$ for the spiral regime, (b)~$y_{\rm th}$ for the funnel regime, (c)~$x_{\rm th}$ for coupled Lorenz oscillators, and (d)~$x_{\rm th}$ for a pair of R\"ossler oscillators in different regimes. The probabilities correspond to the following higher block shift spaces: \mbox{(\textcolor{black}{\hspace{0.1ex}\rule[0.350ex]{0.4cm}{1.5pt}\hspace{0.1ex}}) $\Sigma^{[4]}_{\bm{Z}^{\rm u}}$},  \mbox{(\textcolor{red}{\hspace{0.1ex}\rule[0.350ex]{0.4cm}{1.5pt}\hspace{0.1ex}}) $\Sigma^{[4]}_{\bm{Z}^1}$}, and \mbox{(\textcolor{blue}{\hspace{0.1ex}\rule[0.350ex]{0.4cm}{1.5pt}\hspace{0.1ex}}) $\Sigma^{[4]}_{\bm{Z}^2}$}. The dotted line represents the tolerance $\epsilon{=}3\%$. The curves derive from 30 Monte Carlo simulations for different values of frequency mismatch and initial states. Lines are the median values and the shaded region indicate the second and third quartiles. In (d) two pair of probabilities are presented that correspond to the shift space based on the $x^\prime_{\rm th}$ variable -- black and red -- and on the $y^\prime_{\rm th}$ variable -- orange and blue.}
	\label{fig:threshold}
\end{figure}

\sloppypar{
Following Step~9, we chose $q=4$ and search for a threshold value such that
\mbox{$P\big(\Sigma^{[q]}_{\bm{Z}^{\rm i}}=1\ldots1|z_{\rm th}\big){<}\epsilon$} and \mbox{$P\big(\Sigma^{[q]}_{\bm{Z}^{\rm u}}=1\ldots1|z_{\rm th}\big)$} is largest (close to 1). The reasoning behind this choice of the threshold is that the probability \mbox{$P\big(\Sigma^{[4]}_{\bm{Z}^1} = 1111\big)$}, or \mbox{$P\big(\Sigma^{[4]}_{\bm{Z}^2} = 1111\big)$}, should be small
because the forbidden word is not expected to happen when the oscillators are synchronized and the probability $P\big(\Sigma^{[4]}_{\bm{Z}^{\rm u}} = 1111\big)$ should be much larger than all other probabilities, because the forbidden word is expected to happen as many times as possible in the unsynchronized regime. This is illustrated in
Figure~\ref{fig:threshold}.}

For the case of coupled spiral R\"ossler oscillators, both probabilities $P\big(\Sigma^{[4]}_{\bm{Z}^1} = 1111\big)$ and $P\big(\Sigma^{[4]}_{\bm{Z}^2} = 1111\big)$ converge to zero whilst $P\big(\Sigma^{[4]}_{\bm{Z}^{\rm u}} = 1111\big)$ has median close to one. In this situation $\epsilon\approx0$ was used and the threshold was chosen
soon after the convergence point: $x^\prime_{\rm th} = -16.5$.

As for the coupled funnel oscillators, the probabilities $P\big(\Sigma^{[4]}_{\bm{Z}^1} = 1111\big)$ and $P\big(\Sigma^{[4]}_{\bm{Z}^2} = 1111\big)$ only converge to zero for small values of $P\big(\Sigma^{[4]}_{\bm{Z}^{\rm u}} = 1111\big)$. 
Hence a tolerance of $\epsilon{=}0.03$ was used and the threshold value of $y^\prime_{\rm th}=-18.0$ was chosen.
In this case the probability of finding the forbidden word for the unsynchronized scenario is very high whereas
it is less than 3\% in the synchronization case, which is likely to be good quality phase synchronization.

The same reasoning was used in the case of the coupled Lorenz oscillators. In this case a slightly larger
tolerance was used: $\epsilon{=}0.05$ and the selected threshold was $x^\prime_{\rm th} = 28$.

Determining the partition for the (R\"ossler) spiral-funnel coupling scheme is a bit more intricate as
it would be expected because the dynamics are quite different. The two hypothetical latent spaces that would represent a complete synchronization regime have different topologies and are best described in the 1D first return map by different variables. Figure~\ref{fig:threshold}(d) shows two pair of probabilities corresponding to the shift spaces 
$\left(\Sigma^{[4],x}_{\bm{Z}^{\rm u}},\,\Sigma^{[4]}_{\bm{Z}^{1}}\right)$ -- black and red --
and $\left(\Sigma^{[4],y}_{\bm{Z}^{\rm u}},\,\Sigma^{[4]}_{\bm{Z}^{2}}\right)$ -- orange and blue --, 
which consider threshold points in different variables, $x^\prime_{\rm th}$ and $y^\prime_{\rm th}$, respectively. The pair of probabilities involving $\left(\Sigma^{[4],x}_{\bm{Z}^{\rm u}},\,\Sigma^{[4]}_{\bm{Z}^{1}}\right)$ fits the most the criterion of \mbox{$P\big(\Sigma^{[4]}_{\bm{Z}^{\rm u}} = 1111\big)\approx 1$}, thus the 1D first return map $P:x^\prime[k{-}1]\mapsto x^\prime[k]$ is chosen as source of symbol sequences.
The selected threshold point is $x^\prime_{\rm th}=-18.8$ for the spiral-funnel coupling scheme with $\epsilon=0.05$.

From the narrow interquartiles (Fig.\,\ref{fig:threshold}) in all the examples that use the R\"ossler oscillator it is clear that
the computed word probabilities are not sensitive to initial conditions and to frequency mismatch. This is
not the case for the Lorenz oscillator. Also, when the oscillators are similar there is no noticeable 
difference in using $\bm{Z}^{\rm i}{=}\bm{Z}^1$ or $\bm{Z}^{\rm i}{=}\bm{Z}^2$.

\subsection{Statistics and validation of symbol sequences}

\begin{figure*}[t!]
	\centering
	\includegraphics[width=\linewidth]{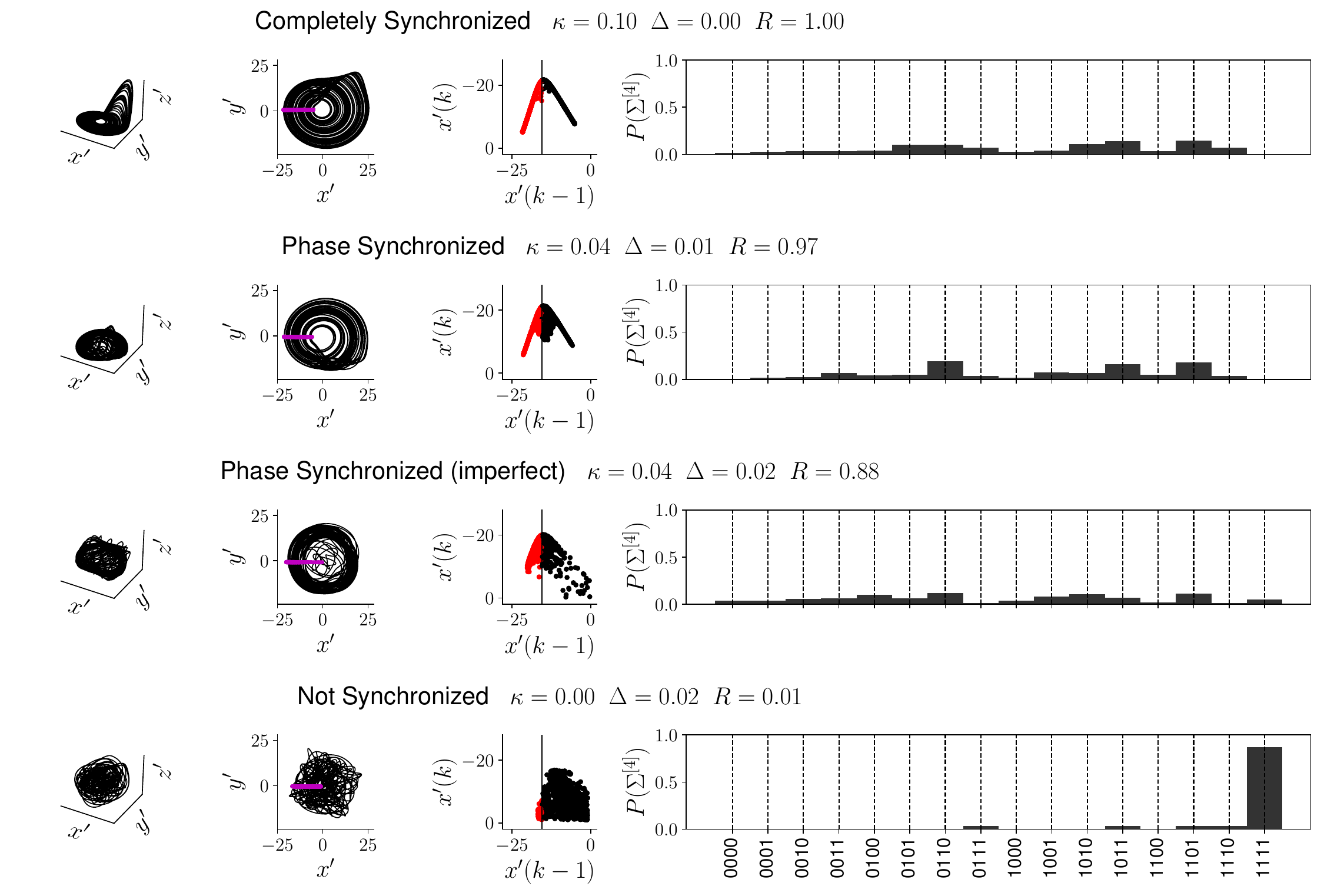}
	\caption{Latent spaces for different levels of synchronization between a pair of \textit{spiral} R\"ossler oscillators. From left to right: 3D latent space, projection of latent space onto the plane $x^\prime\times y^\prime$, first-return map \mbox{$P:x^\prime[k{-}1]\mapsto x^\prime[k]$}, and symbol sequence statistics for the higher block shift space $\Sigma^{[4]}$.}
	\label{fig:spiral}
\end{figure*}

When the quality of phase synchronization is reduced the latent space becomes increasingly disorganized, as
seen in Fig.\,\ref{fig:spiral}.

\begin{figure*}[t!]
	\centering
	\includegraphics[width=\linewidth]{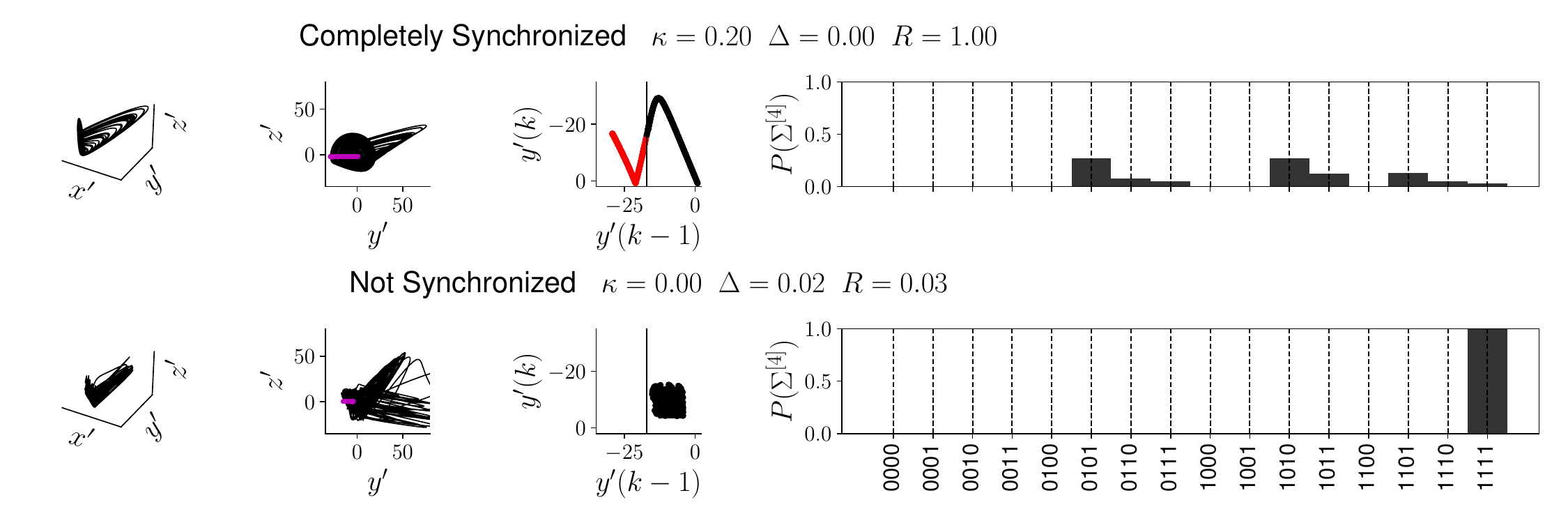}
	\caption{Latent spaces for the cases of complete synchronization and no synchronization between a pair of \textit{funnel} R\"ossler oscillators.  From left to right: 3D latent space, projection of latent space onto the plane $y^\prime\times z^\prime$, first-return map \mbox{$P:y^\prime[k{-}1]\mapsto y^\prime[k]$}, and symbol sequence statistics for the higher block shift space $\Sigma^{[4]}$.}
	\label{fig:funnel}
\end{figure*}

In the completely synchronized regime, the latent space represents the dynamics of an isolated oscillator, which is easily seen by the $x'\times y'$ and $y'\times z'$ projections of the spiral-spiral (Fig.\,\ref{fig:spiral}) and 
funnel-funnel (Fig.\,\ref{fig:funnel}) latent spaces, respectively, and the corresponding characteristic first return maps.

\begin{figure*}[t!]
	\centering
	\includegraphics[width=\linewidth]{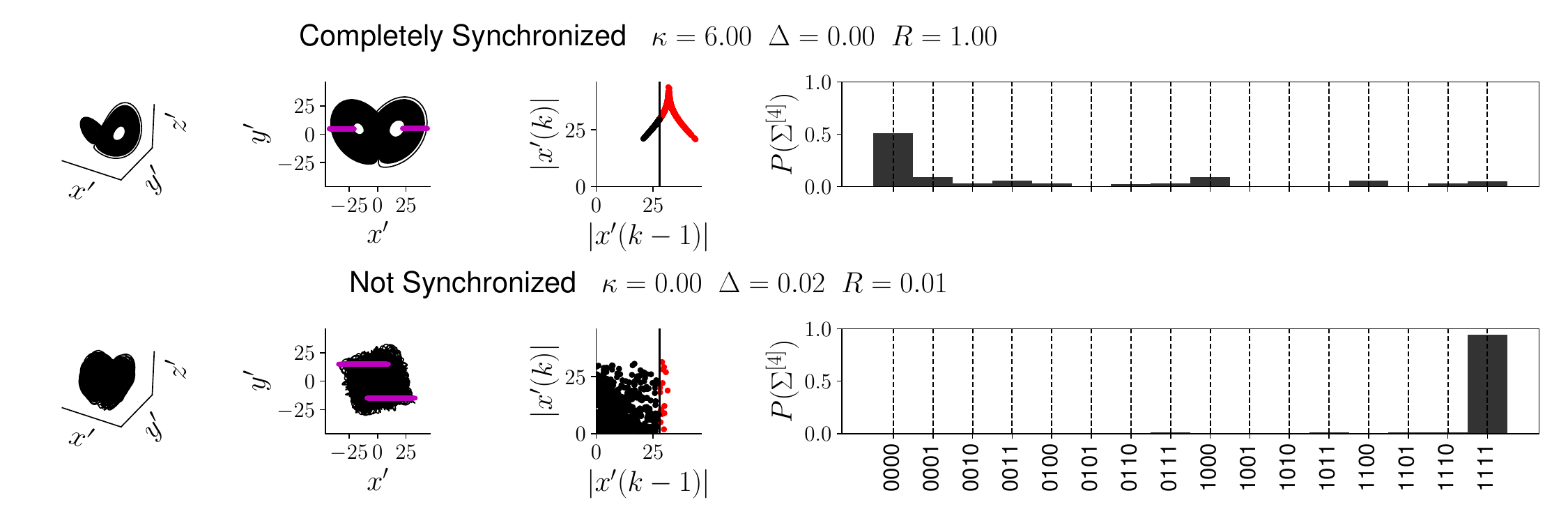}
	\caption{Latent spaces for the cases of complete synchronization and no synchronization between a pair of Lorenz oscillators.}
	\label{fig:lorenz}
\end{figure*}

The resulting symbol sequence statistics shows that no forbidden word appears in the spiral-spiral scheme, whilst 
it does in the funnel-funnel scheme, although with low probability ($\epsilon=0.03$), as expected from the proposed partitioning rule. The same applies for the Lorenz-Lorenz case, where $\epsilon=0.05$ (Fig.\,\ref{fig:lorenz}).

Figure~\ref{fig:spiral} shows two different phase synchronization scenarios: one with good quality 
($\kappa{=}0.038,\Delta{=}0.01$), and one which is imperfect ($\kappa{=}0.038,\Delta{=}0.02$). 
The former presents \textit{phase slips}, which is characterized by a fast increasing of $2\pi$ in the phase difference between the oscillators. Notice that, in the good quality phase synchronization, the corresponding first return map spreads over the interval as if it were {locally} contaminated with noise. In this situation, there is still no forbidden word appearing in the symbol sequence statistics. As the phase synchronization deteriorates and phase slips start to occur, the corresponding latent spaces become increasingly disorganized and the trajectories tend to be drawn to the center of rotation (latent fixed point), and the forbidden word starts to appear. Notice the first return maps and how the trajectories approach the origin of the latent space, which is close to the latent fixed  point, as phase synchronization deteriorates. Finally, in the unsynchronized regime (uncoupled systems), whose latent space is a messy entanglement, it is impossible to recognize the \textit{stretch-and-fold} mechanism that produces the chaotic dynamics in the R\"ossler oscillator. In the latter situation the trajectories visit the region of the fixed point in such a way that the dynamical behavior is practically characterized by a single word (the forbidden word).

We highlight the case of good quality phase synchronization between spiral R\"ossler oscillators (second row of Figure~\ref{fig:spiral}), where the direction of the flow in the latent space is inverted in comparison with the flow of an isolated oscillator. It justifies the proposed section in which the trajectories are sampled in both crossing directions. Furthermore, the case of imperfect phase synchronization (third row of Figure~\ref{fig:spiral}) deserves a deeper examination. 

\begin{figure}[t]
	\centering
	\includegraphics[width=0.5\linewidth]{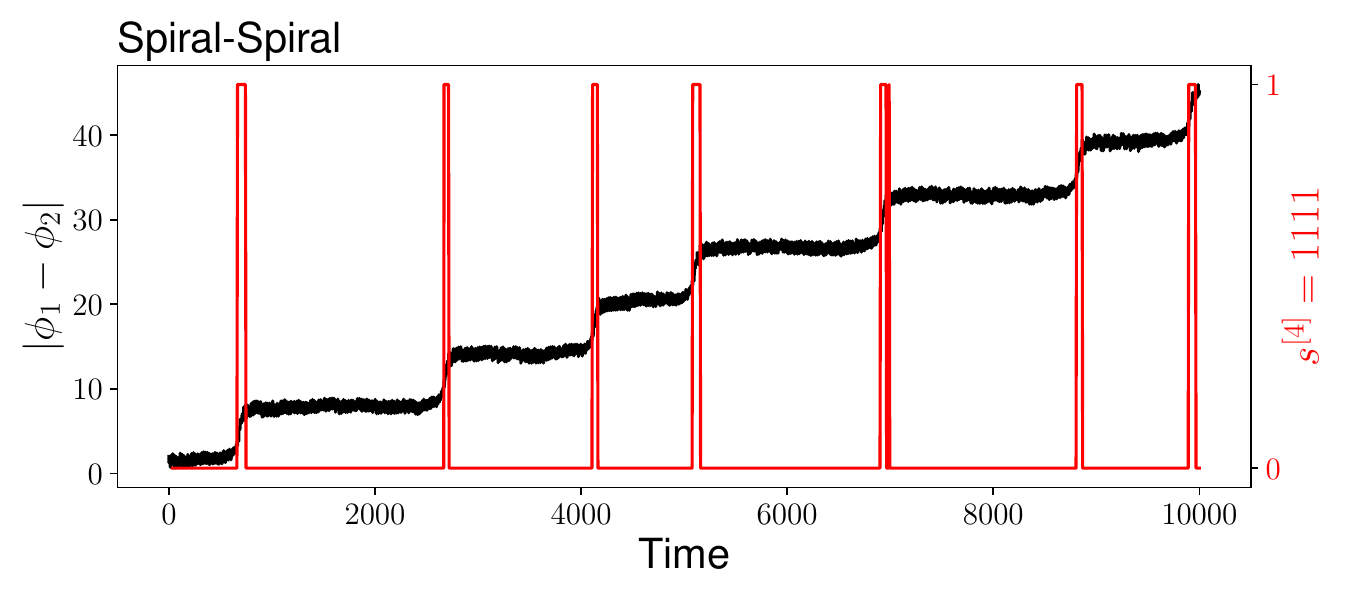}
	\caption{Forbidden word occurrences during imperfect phase synchronization between a pair of spiral R\"ossler oscillators. The figure shows the absolute phase difference between the oscillators \mbox{(\textcolor{black}{\hspace{0.1ex}\rule[0.350ex]{0.4cm}{1.5pt}\hspace{0.1ex}})} and a Boolean time series that indicates the presence of the forbidden word $1111$ \mbox{(\textcolor{red}{\hspace{0.1ex}\rule[0.350ex]{0.4cm}{1.5pt}\hspace{0.1ex}})}.}
	\label{fig:slips}
\end{figure}

Figure~\ref{fig:slips} presents a time series of the absolute phase difference between spiral R\"ossler oscillators during imperfect phase synchronization and a Boolean time series that indicates the presence of the forbidden word.
The effectiveness of the partitioning rule can be verified by noticing that the occurrence of the forbidden word is conditioned to the occurrence of phase slips. This phenomenon does not happen during the observation of phase slips in the funnel-funnel scheme (not shown). We conjecture that it could be a consequence of the use of the tolerance $\epsilon$.

Figure~\ref{fig:spiralFunnel} shows the latent space drawn from the spiral-funnel coupling scheme in two situations: (i) phase synchronized and (ii) unsynchronized (uncoupled). In the first situation, the topology in the latent space is similar to the one from funnel R\"ossler oscillator. It means that the funnel dynamics dominates the spiral one in the latent space. Nonetheless, the spiral dynamical information is not entirely discarded by the PCA. This can be
seen by noticing that the first return map is built using $x^\prime$ variable instead of $y^\prime$.
	
\begin{figure}[t]
	\centering
	\includegraphics[width=0.5\linewidth]{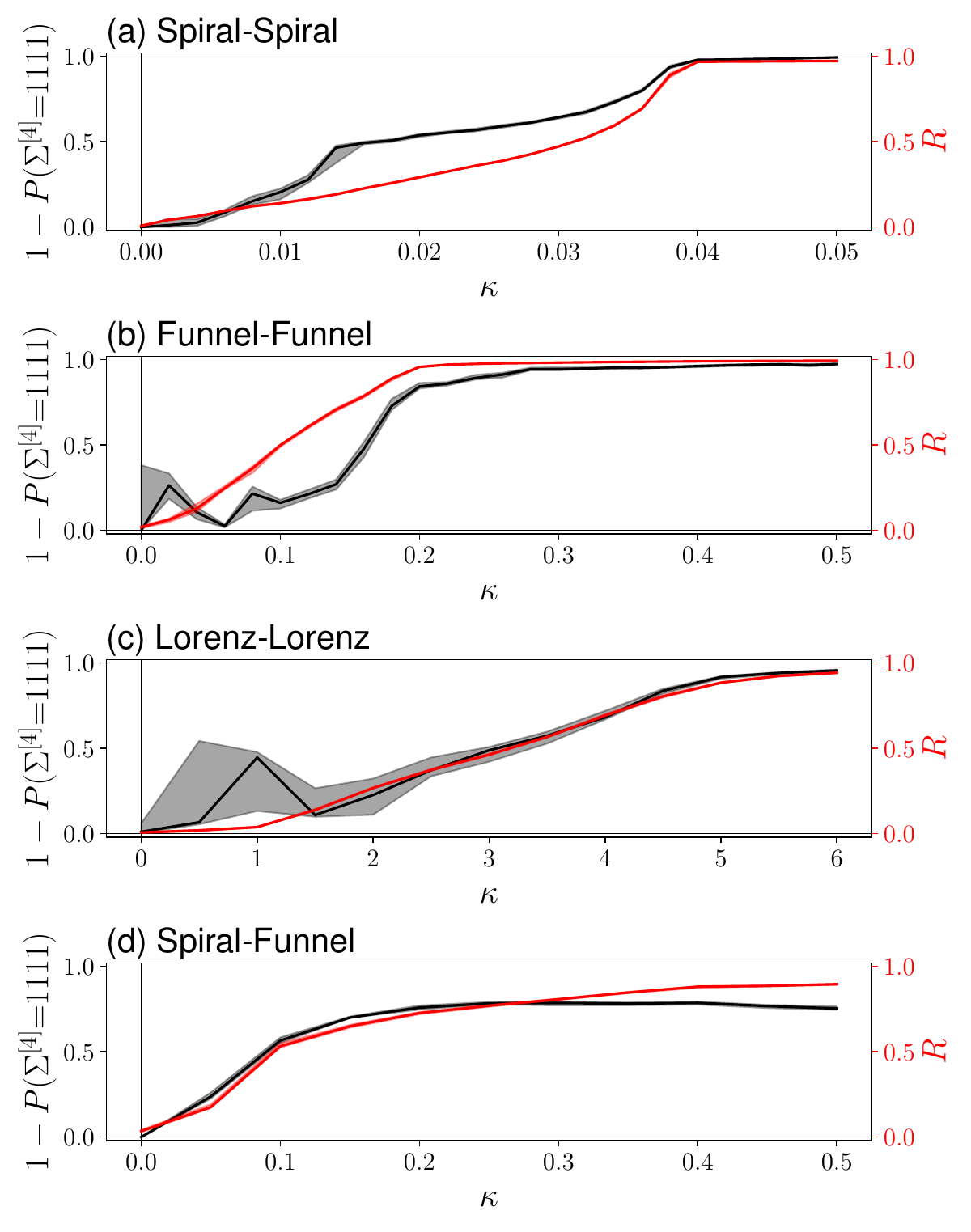}
	\caption{The complement of forbidden word probabilities \mbox{(\textcolor{black}{\hspace{0.1ex}\rule[0.350ex]{0.4cm}{1.5pt}\hspace{0.1ex}})} vs. coupling strength, in comparison with mean phase coherence \mbox{(\textcolor{red}{\hspace{0.1ex}\rule[0.350ex]{0.4cm}{1.5pt}\hspace{0.1ex}})}.}
	\label{fig:coherence}
\end{figure}

Finally, the probability of forbidden word occurrences are compared with the mean phase coherence defined in Eq.~\eqref{eq:meanPhaseCohe}. This is shown in Figure~\ref{fig:coherence}, where we take the complement of the forbidden word probability
\begin{equation}
	\label{eq:complement}
	 \overline{P\big(\Sigma^{[4]}_{\bm{Z}^u} = 1111\big)} = 1{-}P\big(\Sigma^{[4]}_{\bm{Z}^u} = 1111\big).
\end{equation}
Notice that the values of $R$ and of $\overline{P\big(\Sigma^{[4]}_{\bm{Z}^{\rm u}} = 1111\big)}$ have similar general behavior although the latter does not require any phase definition. 
Thus, the complement of the forbidden word probability may be used as a measure of phase synchronization quality. Obviously, the shape of $\overline{P\big(\Sigma^{[4]}_{\bm{Z}^u} = 1111\big)}$ is highly dependent of the chosen partition.

\section{\label{sec:conclusion}Conclusion}

The latent variable subspace has been investigated in the context of phase synchronization analysis. 
It was shown that the orbits in the latent space become disorganized as the phase synchronization onset is weakened. 
Such phenomenon is used to formulate a method based on symbol sequence statistics that yields some measure of phase synchronization quality without a proper definition of phase and of generating partitions. Further studies will provide more insights on how to choose the most convenient non-generating partition. Moreover, the PCA has been sufficient to characterize phase discrepancies between coupled chaotic oscillators. Its advantage over M-SSA is that much less computational power would be required when analyzing more complex networks. In this paper, all state variables of the oscillators are measured. The role of observability and embedding techniques -- when only one variable is measured -- on the latent space topology during phase synchronization onset is a work in progress. The use of a 3D latent space to successfully analyze dynamics on a 6D space suggest that the latent space could be useful in other applications where a specific representation on a lower-dimensional space is called for.

\begin{acknowledgments}
The authors gratefully acknowledge financial support from Programa de P\'os-Graduação em Engenharia Elétrica (PPGEE), CAPES/PROEX Code 001 (HCC) and CNPq grant No.
303412/2019-4 (LAA).
\end{acknowledgments}


\bibliography{../../bib/mychaos,../../bib/clet,../../bib/chaos,../../bib/synch,../../bib/book_cao,aipArticle,synch}


\end{document}